\catcode`\@=11
\expandafter\ifx\csname @iasmacros\endcsname\relax
	\global\let\@iasmacros=\par
\else	\immediate\write16{}
	\immediate\write16{Warning:}
	\immediate\write16{You have tried to input iasmacros more than once.}
	\immediate\write16{}
	\endinput
\fi
\catcode`\@=12


\def\rmb{\seventeenrm}

\def\singlespace{\baselineskip=\normalbaselineskip}
\def\halfspace{\baselineskip=1.5\normalbaselineskip}
\def\doublespace{\baselineskip=2\normalbaselineskip}


\def\AB{\bigskip\parindent=40pt
        \centerline{\bf ABSTRACT}\medskip\halfspace\narrower}
\def\AE{\bigskip\nonarrower\doublespace}
\def\nonarrower{\advance\leftskip by-\parindent
	\advance\rightskip by-\parindent}


\def\boxit#1{\vbox{\hrule\hbox{\vrule\kern3pt
	\vbox{\kern3pt#1\kern3pt}\kern3pt\vrule}\hrule}}

\def\hence{\leavevmode\hbox{\bf .\raise5.5pt\hbox{.}.} }

\def\dalemb#1#2{{\vbox{\hrule height.#2pt
	\hbox{\vrule width.#2pt height#1pt \kern#1pt \vrule width.#2pt}
	\hrule height.#2pt}}}
\def\gtorder{\mathrel{\raise.3ex\hbox{$>$}\mkern-14mu
             \lower0.6ex\hbox{$\sim$}}}
\def\ltorder{\mathrel{\raise.3ex\hbox{$<$}\mkern-14mu
             \lower0.6ex\hbox{$\sim$}}}

\newdimen\fullhsize
\newbox\leftcolumn
\def\twoup{\hoffset=-.5in \voffset=-.25in
  \hsize=4.75in \fullhsize=10in \vsize=6.9in
  \def\fullline{\hbox to\fullhsize}
  \let\lr=L
  \output={\if L\lr
        \global\setbox\leftcolumn=\columnbox\global\let\lr=R \advancepageno
      \else \doubleformat \global\let\lr=L\fi
    \ifnum\outputpenalty>-20000 \else\dosupereject\fi}
  \def\doubleformat{\shipout\vbox{
    \fullline{\box\leftcolumn\hfil\columnbox}\advancepageno}}
  \def\columnbox{\leftline{\vbox{\makeheadline\pagebody\makefootline}}}
  \tolerance=1000 }
\catcode`\@=11					



\font\fiverm=cmr5				
\font\fivemi=cmmi5				
\font\fivesy=cmsy5				
\font\fivebf=cmbx5				

\skewchar\fivemi='177
\skewchar\fivesy='60


\font\sixrm=cmr6				
\font\sixi=cmmi6				
\font\sixsy=cmsy6				
\font\sixbf=cmbx6				

\skewchar\sixi='177
\skewchar\sixsy='60


\font\sevenrm=cmr7				
\font\seveni=cmmi7				
\font\sevensy=cmsy7				
\font\sevenit=cmti7				
\font\sevenbf=cmbx7				

\skewchar\seveni='177
\skewchar\sevensy='60


\font\eightrm=cmr8				
\font\eighti=cmmi8				
\font\eightsy=cmsy8				
\font\eightit=cmti8				
\font\eightbf=cmbx8				

\skewchar\eighti='177
\skewchar\eightsy='60


\font\ninei=cmmi9
\font\ninesy=cmsy9

\skewchar\ninei='177
\skewchar\ninesy='60


\font\tenrm=cmr10				
\font\teni=cmmi10				
\font\tensy=cmsy10				
\font\tenex=cmex10				
\font\tenit=cmti10				
\font\tensl=cmsl10				
\font\tenbf=cmbx10				
\font\tentt=cmtt10				
\font\tenss=cmss10				
\font\tensc=cmcsc10				
\font\tenbi=cmmib10				

\skewchar\teni='177
\skewchar\tenbi='177
\skewchar\tensy='60

\def\tenpoint{\ifmmode\err@badsizechange\else
	\textfont0=\tenrm \scriptfont0=\sevenrm \scriptscriptfont0=\fiverm
	\textfont1=\teni  \scriptfont1=\seveni  \scriptscriptfont1=\fivemi
	\textfont2=\tensy \scriptfont2=\sevensy \scriptscriptfont2=\fivesy
	\textfont3=\tenex \scriptfont3=\tenex   \scriptscriptfont3=\tenex
	\textfont4=\tenit \scriptfont4=\sevenit \scriptscriptfont4=\sevenit
	\textfont5=\tensl
	\textfont6=\tenbf \scriptfont6=\sevenbf \scriptscriptfont6=\fivebf
	\textfont7=\tentt
	\textfont8=\tenbi \scriptfont8=\seveni  \scriptscriptfont8=\fivemi
	\def\rm{\tenrm\fam=0 }%
	\def\it{\tenit\fam=4 }%
	\def\sl{\tensl\fam=5 }%
	\def\bf{\tenbf\fam=6 }%
	\def\tt{\tentt\fam=7 }%
	\def\ss{\tenss}%
	\def\sc{\tensc}%
	\def\bmit{\fam=8 }%
	\rm\setparameters\setbaselines\fi}


\font\twelverm=cmr12				
\font\twelvei=cmmi12				
\font\twelvesy=cmsy10	scaled\magstep1		
\font\twelveex=cmex10	scaled\magstep1		
\font\twelveit=cmti12				
\font\twelvesl=cmsl12				
\font\twelvebf=cmbx12				
\font\twelvett=cmtt12				
\font\twelvess=cmss12				
\font\twelvesc=cmcsc10	scaled\magstep1		
\font\twelvebi=cmmib10	scaled\magstep1		

\skewchar\twelvei='177
\skewchar\twelvebi='177
\skewchar\twelvesy='60

\def\twelvepoint{\ifmmode\err@badsizechange\else
	\textfont0=\twelverm \scriptfont0=\eightrm \scriptscriptfont0=\sixrm
	\textfont1=\twelvei  \scriptfont1=\eighti  \scriptscriptfont1=\sixi
	\textfont2=\twelvesy \scriptfont2=\eightsy \scriptscriptfont2=\sixsy
	\textfont3=\twelveex \scriptfont3=\tenex   \scriptscriptfont3=\tenex
	\textfont4=\twelveit \scriptfont4=\eightit \scriptscriptfont4=\sevenit
	\textfont5=\twelvesl
	\textfont6=\twelvebf \scriptfont6=\eightbf \scriptscriptfont6=\sixbf
	\textfont7=\twelvett
	\textfont8=\twelvebi \scriptfont8=\eighti  \scriptscriptfont8=\sixi
	\def\rm{\twelverm\fam=0 }%
	\def\it{\twelveit\fam=4 }%
	\def\sl{\twelvesl\fam=5 }%
	\def\bf{\twelvebf\fam=6 }%
	\def\tt{\twelvett\fam=7 }%
	\def\ss{\twelvess}%
	\def\sc{\twelvesc}%
	\def\bmit{\fam=8 }%
	\rm\setparameters\setbaselines\fi}


\font\fourteenrm=cmr12	scaled\magstep1		
\font\fourteeni=cmmi12	scaled\magstep1		
\font\fourteensy=cmsy10	scaled\magstep2		
\font\fourteenex=cmex10	scaled\magstep2		
\font\fourteenit=cmti12	scaled\magstep1		
\font\fourteensl=cmsl12	scaled\magstep1		
\font\fourteenbf=cmbx12	scaled\magstep1		
\font\fourteentt=cmtt12	scaled\magstep1		
\font\fourteenss=cmss12	scaled\magstep1		
\font\fourteensc=cmcsc10 scaled\magstep2	
\font\fourteenbi=cmmib10 scaled\magstep2	

\skewchar\fourteeni='177
\skewchar\fourteenbi='177
\skewchar\fourteensy='60

\def\fourteenpoint{\ifmmode\err@badsizechange\else
	\textfont0=\fourteenrm \scriptfont0=\tenrm \scriptscriptfont0=\sevenrm
	\textfont1=\fourteeni  \scriptfont1=\teni  \scriptscriptfont1=\seveni
	\textfont2=\fourteensy \scriptfont2=\tensy \scriptscriptfont2=\sevensy
	\textfont3=\fourteenex \scriptfont3=\tenex \scriptscriptfont3=\tenex
	\textfont4=\fourteenit \scriptfont4=\tenit \scriptscriptfont4=\sevenit
	\textfont5=\fourteensl
	\textfont6=\fourteenbf \scriptfont6=\tenbf \scriptscriptfont6=\sevenbf
	\textfont7=\fourteentt
	\textfont8=\fourteenbi \scriptfont8=\tenbi \scriptscriptfont8=\seveni
	\def\rm{\fourteenrm\fam=0 }%
	\def\it{\fourteenit\fam=4 }%
	\def\sl{\fourteensl\fam=5 }%
	\def\bf{\fourteenbf\fam=6 }%
	\def\tt{\fourteentt\fam=7}%
	\def\ss{\fourteenss}%
	\def\sc{\fourteensc}%
	\def\bmit{\fam=8 }%
	\rm\setparameters\setbaselines\fi}


\font\seventeenrm=cmr10 scaled\magstep3		


\newdimen\rp@
\newcount\@basestretchnum
\newskip\@baseskip
\newskip\headskip
\newskip\footskip


\def\setparameters{\rp@=.1em
	\headskip=24\rp@
	\footskip=\headskip
	\delimitershortfall=5\rp@
	\nulldelimiterspace=1.2\rp@
	\scriptspace=0.5\rp@
	\abovedisplayskip=10\rp@ plus3\rp@ minus5\rp@
	\belowdisplayskip=10\rp@ plus3\rp@ minus5\rp@
	\abovedisplayshortskip=5\rp@ plus2\rp@ minus4\rp@
	\belowdisplayshortskip=10\rp@ plus3\rp@ minus5\rp@
	\normallineskip=\rp@
	\lineskip=\normallineskip
	\normallineskiplimit=0pt
	\lineskiplimit=\normallineskiplimit
	\jot=3\rp@
	\setbox0=\hbox{\the\textfont3 B}\p@renwd=\wd0
	\skip\footins=12\rp@ plus3\rp@ minus3\rp@
	\skip\topins=0pt plus0pt minus0pt}


\def\setbaselines{\maxdepth=4\rp@\baselinestretch=\@basestretchnum}


\def\baselinestretch{\afterassignment\@basestretch\@basestretchnum}
\def\@basestretch{%
	\@baseskip=12\rp@ \divide\@baseskip by1000
	\normalbaselineskip=\@basestretchnum\@baseskip
	\baselineskip=\normalbaselineskip
	\bigskipamount=\the\baselineskip
		plus.25\baselineskip minus.25\baselineskip
	\medskipamount=.5\baselineskip
		plus.125\baselineskip minus.125\baselineskip
	\smallskipamount=.25\baselineskip
		plus.0625\baselineskip minus.0625\baselineskip
	\setbox\strutbox=\hbox{\vrule height.708\baselineskip
		depth.292\baselineskip width0pt }}



\def\makeheadline{\vbox to0pt{\baselinestretch=1000
	\vskip-\headskip \vskip1.5pt
	\line{\vbox to\ht\strutbox{}\the\headline}\vss}\nointerlineskip}

\def\makefootline{\baselineskip=\footskip\line{\the\footline}}

\def\big#1{{\hbox{$\left#1\vbox to8.5\rp@ {}\right.\n@space$}}}
\def\Big#1{{\hbox{$\left#1\vbox to11.5\rp@ {}\right.\n@space$}}}
\def\bigg#1{{\hbox{$\left#1\vbox to14.5\rp@ {}\right.\n@space$}}}
\def\Bigg#1{{\hbox{$\left#1\vbox to17.5\rp@ {}\right.\n@space$}}}


\mathchardef\alpha="710B
\mathchardef\beta="710C
\mathchardef\gamma="710D
\mathchardef\delta="710E
\mathchardef\epsilon="710F
\mathchardef\zeta="7110
\mathchardef\eta="7111
\mathchardef\theta="7112
\mathchardef\iota="7113
\mathchardef\kappa="7114
\mathchardef\lambda="7115
\mathchardef\mu="7116
\mathchardef\nu="7117
\mathchardef\xi="7118
\mathchardef\pi="7119
\mathchardef\rho="711A
\mathchardef\sigma="711B
\mathchardef\tau="711C
\mathchardef\upsilon="711D
\mathchardef\phi="711E
\mathchardef\chi="711F
\mathchardef\psi="7120
\mathchardef\omega="7121
\mathchardef\varepsilon="7122
\mathchardef\vartheta="7123
\mathchardef\varpi="7124
\mathchardef\varrho="7125
\mathchardef\varsigma="7126
\mathchardef\varphi="7127
\mathchardef\imath="717B
\mathchardef\jmath="717C
\mathchardef\ell="7160
\mathchardef\wp="717D
\mathchardef\partial="7140
\mathchardef\flat="715B
\mathchardef\natural="715C
\mathchardef\sharp="715D


\def\err@badsizechange{%
	\immediate\write16{--> Size change not allowed in math mode, ignored}}

\baselinestretch=1000
\tenpoint

\catcode`\@=12					

\twelvepoint
\doublespace
{\nopagenumbers{
\rightline{IASSNS-HEP-97/15}
\rightline{~~~March, 1997}
\bigskip\bigskip
\centerline{\rmb The Matrix Model for M Theory as an Exemplar}
\centerline{\rmb of Trace (or Generalized Quantum) Dynamics}
\medskip
\centerline{\it Stephen L. Adler
}
\centerline{\bf Institute for Advanced Study}
\centerline{\bf Princeton, NJ 08540}
\medskip
\bigskip\bigskip
\leftline{\it Send correspondence to:}
\medskip
{\singlespace\leftline{Stephen L. Adler}
\leftline{Institute for Advanced Study}
\leftline{Olden Lane, Princeton, NJ 08540}
\leftline{Phone 609-734-8051; FAX 609-924-8399; email adler@ias.
edu}}
\bigskip\bigskip
}}
\vfill\eject
\pageno=2
\AB
We show that the recently proposed matrix model for M theory obeys the 
cyclic trace assumptions underlying generalized quantum or trace dynamics.  
This permits a verification of supersymmetry as an operator calculation, 
and a calculation of the supercharge density algebra by using the generalized 
Poisson bracket, in a basis-independent manner that makes no reference to  
individual matrix elements.  Implications for quantization of the model 
are discussed.  Our results are a special case of a general 
result presented elsewhere, that all rigid supersymmetry theories can be 
extended to give supersymmetric trace dynamics theories, in which the 
supersymmetry algebra is represented by the generalized Poisson bracket 
of trace supercharges, constructed from fields that form 
a noncommutative trace class graded operator algebra.  
\AE
\bigskip\bigskip
\vfill\eject
\pageno=3

Recently Banks, Fischler, Shenker, and Susskind  
[1] have suggested a supersymmetric matrix model 
(based on earlier supersymmetric quantum mechanics models [2]) 
for use as a tool to study uncompactified eleven dimensional $M$ theory, 
and more recently Banks, Seiberg, and Shenker [3] have computed the 
supersymmetry charge density algebra in this model.  Although they 
performed this computation using the Poisson brackets for the 
individual matrix elements, the fact that the final answers are expressible 
directly in matrix terms suggests that it should be possible to perform 
the calculation using matrix or operator methods throughout.  We shall 
show in this paper that the operator dynamics that we have proposed [4] 
and studied with various collaborators [5, 6] is admirably suited for this 
purpose.  In the original papers we termed this operator dynamics 
``generalized quantum dynamics'', but A. Millard in his thesis [7] uses the 
briefer and more descriptive name ``trace dynamics'', which we shall use 
henceforth.  What we shall show here is that trace dynamics permits an 
operator treatment of the matrix model for M theory, and generalizes this 
model to the case in which the matrix elements are themselves noncommutative 
matrices (as would be of interest for renormalization group block spin 
schemes).  More generally, we shall show elsewhere [8] that all  rigid 
supersymmetry theories can be extended to supersymmetric trace dynamics 
theories, giving supersymmetry representations over trace class 
noncommutative graded operator algebras. Because 
of the current interest in the model of [1], 
and the fact that this model does not require the gauge fixing technicalities
that enter into the case of supersymmetric Yang-Mills theory, it is well  
suited to a brief account illustrative of our general approach.  

Let $X_1$ and $X_2$ be two $N \times N$ matrices with complex (or more 
generally, complex Grassmann even) matrix elements, 
and Tr the ordinary matrix trace, which obeys the cyclic property 
$${\rm Tr} X_1 X_2 = \sum_{m,n} (X_1)_{mn} (X_2)_{nm}
=\sum_{m,n} (X_2)_{nm}(X_1)_{mn} = {\rm Tr} X_2 X_1~~~.\eqno(1a)$$  
Correspondingly, let $\theta_1$ and $\theta_2$ be two $N \times N$ matrices 
with complex Grassmann odd matrix elements, which anticommute rather than 
commute, so that the cyclic property for these takes the form 
$${\rm Tr} \theta_1 \theta_2 =\sum_{m,n} (\theta_1)_{mn} (\theta_2)_{nm}
=-\sum_{m,n} (\theta_2)_{nm} (\theta_1)_{mn} = -{\rm Tr} \theta_2 \theta_1
~~~.\eqno(1b)$$
The cyclic properties of Eqs.~(1a, 1b) are just those assumed for the trace 
operation {\bf Tr} of trace dynamics (although in Refs. [4-6] 
the fermionic operators  
are realized as matrices with  complex matrix elements, all of 
which anticommute with a grading operator $(-1)^F$); we shall continue 
here to use the notation Tr, with the understanding that fermionic matrices 
are Grassmann odd matrices obeying Eq.~(1b), while bosonic matrices are 
Grassmann even matrices obeying  
Eq.~(1a).  From Eqs.~(1a) and (1b), one immediately derives the trilinear 
cyclic identities 
$$\eqalign{
{\rm Tr} X_1[X_2,X_3]=&{\rm Tr}X_2[X_3,X_1]={\rm Tr}X_3[X_1,X_2] \cr
{\rm Tr} X_1\{X_2,X_3\}=&{\rm Tr}X_2\{X_3,X_1\}={\rm Tr}X_3\{X_1,X_2\} \cr
{\rm Tr} X\{\theta_1,\theta_2\}=&{\rm Tr}\theta_1[\theta_2,X]=
{\rm Tr}\theta_2[\theta_1,X] \cr
{\rm Tr}\theta_1\{X,\theta_2\}=&{\rm Tr}\{\theta_1,X\}\theta_2=
{\rm Tr}[\theta_1,\theta_2]X ~~~,\cr
}\eqno(1c)$$
which are repeatedly used below.  

The basic observation of trace dynamics is that given the trace of a 
polynomial $P$ constructed from noncommuting operator variables, one 
can define a derivative of the number $ {\rm Tr} P$ 
with respect to an 
operator variable $\cal O$ by varying and then cyclically permuting so that  
in each term the factor $\delta {\cal O}$ stands on the right, 
giving the fundamental definition 
$$\delta {\rm Tr} P ={\rm Tr}  {\delta {\rm Tr} P \over \delta {\cal O} }
\delta {\cal O} ~~~,\eqno(2a)$$
or in the condensed notation with ${\bf P} \equiv {\rm Tr} P$,  
$$\delta {\bf P} ={\rm Tr}  {\delta {\bf P} \over \delta {\cal O} }
\delta {\cal O} ~~~.\eqno(2b)$$
In Refs.~[4, 5] it is shown that using this definition, one can construct 
a complete Lagrangian and Hamiltonian dynamics for systems with noncommuting 
graded operator variables and a trace Lagrangian, in which 
a generalized Poisson bracket plays the role played 
in classical mechanics by the classical 
Poisson bracket, or by the commutator in quantum mechanics.  In Ref.~[6] 
it is further shown that this dynamical system has the remarkable 
property that its {\it statistical mechanics} gives complex 
quantum field theory, with 
ensemble averages of the operator variables effectively obeying  
standard canonical commutation relations.  

Let us now turn to the matrix model for $M$ theory.  It is described by 
the trace Lagrangian {\bf L} given by 
$${\bf L}={\rm Tr}  \left( {1\over 2} D_t X_i D_t X^i
+ i \theta^T D_t \theta 
+{1\over 4} [X_i,X_j] [X^i,X^j] + \theta^T \gamma_i [\theta, X^i]\right)~~~,
\eqno(3)$$
with $D_t {\cal O}=\partial_t {\cal O} -i [A_0, {\cal O}]$.  In Eq. (3a), 
a summation convention is understood on the indices $i,j$ which range 
from 1 to 9; $A_0$ and the $X_i$ are self-adjoint $N \times N$  
bosonic matrices with complex number 
matrix elements, while $\theta$ is a 16-component fermionic spinor 
each element of 
which is a self-adjoint $N \times N$ complex Grassmann matrix, 
with the transpose $T$ acting 
only on the spinor structure but not on the $N \times N$ matrices, so that 
$\theta^T$ is simply the 16 component row spinor corresponding to the 
16 component column spinor $\theta$.  The potential $A_0$ has no kinetic 
term and so is a pure gauge degree of freedom.  Finally, the $\gamma_i$ are a 
set 
of nine $16 \times 16$ matrices, which are related to the standard $
32 \times 32$ matrices $\Gamma_{\mu}$ as well as to the Dirac matrices of 
spin(8) as described in Danielsson, Ferretti, and Sundborg [9].  This 
finishes the specification of the model. 

Before turning to a study of the model's dynamics and supersymmetries, let  
us summarize some of the properties of the real, symmetric  
matrices $\gamma_i$ that are 
needed.  These 
matrices satisfy the anticommutator algebra 
$$\{ \gamma_i, \gamma_j \} =2 \delta_{ij} ~~~,\eqno(4a)$$
as well as the cyclic identity (which follows by projection from Eq.~(4.A.6)
of Green, Schwarz, and Witten [10]), 
$$\sum_{{\rm cycle}~~ p \rightarrow q \rightarrow n \rightarrow p} 
(\delta^{mn}\delta^{pq}-
\gamma_i^{mn}\gamma_i^{pq})=0~~~,\eqno(4b)$$
with $i$ again summed over and with the indices $m,n,p,q$ spinorial indices 
ranging from 1 to 16.  Defining 
$$\gamma_{ij}={1 \over 2} [\gamma_i, \gamma_j] ~~~,\eqno(4c)$$
so that 
$$\gamma_i \gamma_j = \delta_{ij}+\gamma_{ij}~~~,\eqno(4d)$$
one readily derives from  Eq.~(4b) an identity given in  Ref.~[3], 
$$\gamma_{ij}^{mn}\gamma_i^{pq}+\gamma_{ij}^{pq}\gamma_i^{mn}
+ (m \leftrightarrow p )=2(\gamma_j^{nq} \delta^{mp}-
\gamma_j^{mp}\delta^{nq})
~~~.\eqno(5a)$$
By standard gamma matrix manipulations using Eq.~(4a), one also derives 
the the fact that the matrix 
$$A_{ijk}=\gamma_i\gamma_j\gamma_k-\delta_{ij}\gamma_k+\delta_{ik}
\gamma_j- \delta_{jk} \gamma_i \eqno(5b)$$
is totally antisymmetric in the indices $i,j,k$ (it is just the 
antisymmetrized product $\gamma_{[i}\gamma_j\gamma_{k]}$ with normalization 
factor ${1\over 6})$, as well as the identity 
$${1\over 2}\{\gamma_{\ell m},\gamma_{ij} \}= \gamma_{[\ell}\gamma_m\gamma_i
\gamma_{j]}
+\delta_{\ell j} \delta_{im}- \delta_{mj}\delta_{i\ell}~~~,\eqno(5c)$$         
   
with the first term on the right the antisymmetrized  
product including normalization factor ${1 \over 24}$.  

Now let us turn to dynamics.  From the trace Lagrangian of Eq.~(3), using the 
definition of Eq.~(2) we find the operator derivatives 
$$\eqalign{
{\delta{\bf L} \over \delta A_0}=&-i[X^i,D_t X_i]-2\theta^T\theta  \cr
{\delta{\bf L} \over \delta X_i}=&-i[D_tX^i,A_0]+[[X^j,X^i],X_j]
+2\theta^T\gamma^i\theta \cr
{\delta{\bf L} \over \delta (\partial_tX_i)}=&D_tX^i   \cr
{\delta{\bf L} \over \delta \theta}=&-iD_t\theta^T+[\theta^T,A_0]
-2[\theta^T\gamma_i,X^i]  \cr
{\delta{\bf L} \over \delta (\partial_t \theta)}=&i\theta^T~~~.\cr
}\eqno(6a)$$
Substituting these into the operator Euler-Lagrange equations [4]
$${\partial \over \partial t} {\delta {\bf L} \over 
\delta (\partial_t {\cal O}) }
={\delta {\bf L} \over \delta {\cal O} }~~~,\eqno(6b)$$
and regrouping terms, we get the equations of motion of the matrix 
model 
$$\eqalign{
D_t^2X^i=&[[X^j,X^i],X_j]+2\theta^T \gamma^i \theta   \cr
D_t\theta^T=&i[\theta^T \gamma_i,X^i]~~ \Rightarrow~~ D_t\theta=
i[\gamma_i \theta, X^i]  ~~~,\cr
}\eqno(7a)$$
together with the constraint
$$\tilde C \equiv [X^i,D_t X_i]-2i\theta^T \theta=0~~~.\eqno(7b)$$
To transform the dynamics to trace Hamiltonian form, we define the 
canonical momenta $p_{X_i}$ and $p_{\theta}$ by 
$$\eqalign{
p_{X_i}=&{\delta {\bf L} \over \delta (\partial_t X_i) }=D_tX^i \cr
p_{\theta}=&{\delta {\bf L} \over \delta (\partial_t \theta) }=i \theta^T
~~~,\cr
}\eqno(8a)$$
so that the trace Hamiltonian is given by 
$${\bf H}={\rm Tr}(p_{X_i}\partial_t X_i + p_{\theta} \partial_t \theta )
-{\bf L}             
={\rm Tr}\left({1 \over 2} p_{X_i} p_{X^i} - {1 \over 4}[X_i,X_j][X^i,X^j]
+i p_{\theta} \gamma_i [\theta,X^i] +i A_0 \tilde C \right)~~~.\eqno(8b)$$
Note that because $p_{\theta}=i\theta^T$, it is necessary to write the 
trace Hamiltonian in a form that is manifestly symmetric under the 
replacements $p_{\theta} \to i \theta^T,~~\theta \to -ip_{\theta}^T$; it is 
easy to check that by virtue of the cyclic identities of Eq.(1c) and the 
symmetry of $\gamma_i$ that Eq.~(8b) has this symmetry.  
The Hamilton equations following from the trace Hamiltonian of Eq.~(8b) 
are [4]
$$\eqalign{
\partial_t X_i =&{\delta {\bf H} \over \delta p_{X_i} }     \cr
\partial_t p_{X_i}=&-{\delta {\bf H} \over \delta X_i }     \cr
\partial_t \theta =&-{\delta {\bf H} \over \delta p_{\theta} } \cr
\partial_t p_{\theta}=&-{\delta {\bf H} \over \delta \theta} ~~~,\cr
}\eqno(9)$$
and it is easy to check that they are the same as the operator Euler-Lagrange 
equations derived above.  

So far we have reproduced standard results of the matrix model, but have 
broken no new ground.  Now let us consider the variation of the trace 
Lagrangian, calculated from 
$$\delta {\bf L} ={\rm Tr}\left({\delta {\bf L} \over \delta A_0}\delta A_0 
 +       {\delta {\bf L} \over \delta (\partial_tX_i)} \delta (\partial_tX_i)
 +       {\delta {\bf L} \over \delta X_i    }  \delta X_i
 +       {\delta {\bf L} \over \delta (\partial_t \theta)} 
                                              \delta (\partial_t \theta)  
 +       {\delta {\bf L} \over \delta \theta}  \delta \theta  \right)
 ~~~,\eqno(10a)$$
for the supersymmetry transformation defined by 
$$\eqalign{
\delta X^i=&-2 \epsilon^T \gamma^i \theta = 2 \theta^T \gamma^i \epsilon \cr
\delta \theta =&-\left(iD_tX^i \gamma_i +{1 \over 2}[X^i,X^j] \gamma_{ij}
\right) \epsilon + \epsilon^{\prime}  \cr
\delta A_0=&-2 \epsilon^T \theta=2 \theta^T \epsilon   ~~~.\cr
}\eqno(10b)$$
Here $\epsilon$ and $\epsilon^{\prime}$ are 16 component Grassmann 
$c$-number spinors, that is, they are column vectors each of whose 16 
components is an independent $1 \times 1$ Grassmann matrix.  Using {\it only} 
the cyclic trace identities and gamma matrix properties given above, it is 
a matter of straightforward but lengthy calculation to verify that the 
trace Lagrangian is invariant under the transformation of Eq.~(10b) 
when $\epsilon$ and $\epsilon^{\prime}$ are time independent.  Note that 
in this calculation the variables $X_i$, $\theta$, and $A_0$ are treated 
simply as noncommuting 
operators that are unspecified apart from their bosonic or fermionic 
character; in particular, we do not have to assume that their matrix 
elements when they are written as $N \times N$ matrices are $c$-numbers 
or Grassmann $c$-numbers.  In other words, we have shown that the matrix 
model of Eq.~(3) is still supersymmetric when viewed as a model 
over general noncommutative trace class graded operator variables.

When $\epsilon$ has a time dependence $\delta {\bf L}$ 
is no longer zero, but instead is given by 
$$\eqalign{
\delta{\bf L}=&\partial_t {\rm Tr} \big[ -i \theta^T \epsilon^{\prime}
+\big( \theta^T \gamma_i D_t X^i + {1 \over 2}i\theta^T \gamma_{ij} [X^i,X^j]
\big) \epsilon \big] \cr    
+&{\rm Tr} \big[ 2i\theta^T\partial_t\epsilon^{\prime} 
+\big( 2 \theta^T\gamma_i D_t X^i  - i \theta^T\gamma_{ij} [X^i,X^j]  \big)
\partial_t \epsilon    \big] ~~~.\cr
}\eqno(11a)$$
This identifies the trace supercharges ${\bf Q}^{\prime}_{\alpha}$ and 
${\bf Q}_{\alpha}$ as
$$\eqalign{
{\bf Q}^{\prime}_{\alpha}=&{\rm Tr}  2i\theta^T \alpha \cr 
{\bf Q}_{\alpha} =&{\rm Tr}\big( 2\theta^T\gamma_iD_tX_i -i \theta^T 
\gamma_{ij} [X^i,X^j] \big) \alpha  ~~~,\cr
}\eqno(11b)$$
and their conservation is easily checked using the equations of motion
and constraint of Eqs.~(8a, b) and the identities of Eqs.~(4, 5).  
To check the supersymmetry algebra, 
we must first write the supercharges of Eq.~(11b) 
in Hamiltonian form, symmetrized with respect to  
$p_{\theta}$ and $i \theta^T$, giving 
$$\eqalign{
{\bf Q}^{\prime}_{\alpha}=& {\rm Tr}(p_{\theta}+i\theta^T) \alpha \cr 
{\bf Q}_{\alpha}=&-{\rm Tr}(p_{\theta}+i\theta^T)\big( i\gamma_ip_{X_i} 
+{1\over 2} \gamma_{ij} [X^i,X^j] \big) \alpha  ~~~.\cr
}\eqno(12a)$$
In addition to the supercharge ${\bf Q}_{\alpha}$, Ref.~[3] also introduces 
a supercharge density; in our language this can be written in terms of  
a Grassmann spinor {\it operator} $\beta$ as 
$${\bf Q}_{\beta}=-{1\over 2}{\rm Tr} \big\{ (p_{\theta}+i\theta^T) ,
i\gamma_ip_{X_i}+{1\over 2}\gamma_{ij}[X^i,X^j] \big\} \beta~~~.\eqno(12b)$$

Let us now introduce the generalized Poisson bracket [4, 5] corresponding 
to the Hamiltonian structure of our model, defined for 
any ${\bf A}={\rm Tr}A$ and ${\bf B}={\rm Tr}B$ by 
$$\{{\bf A},{\bf B} \}={\rm Tr}\left(  {\delta{\bf A} \over \delta X_i}
{\delta{\bf B} \over \delta p_{X_i}}-{ \delta{\bf B} \over \delta X_i}
{\delta{\bf A} \over \delta p_{X_i}} -{\delta{\bf A}\over \delta \theta}
{\delta {\bf B} \over \delta p_{\theta} }+{\delta {\bf B} \over 
\delta \theta}
{\delta {\bf A} \over \delta p_{\theta} } \right)~~~.\eqno(13)$$
We can now use the generalized Poisson bracket to give a basis-independent 
evaluation of the supersymmetry charge algebra.  For the brackets involving 
${\bf Q}^{\prime}$, we easily find 
$$\eqalign{
\{ {\bf Q}^{\prime}_{\alpha} , {\bf Q}^{\prime}_{\beta} \}=&
-2i {\rm Tr} \alpha \beta  \cr
\{ {\bf Q}^{\prime}_{\alpha} , {\bf Q}_{\beta} \} =&
2i {\rm Tr} \big(i \gamma_i^{ab}p_{X_i} +{1\over 2}\gamma_{ij}^{ab}
[X^i,X^j] \big) \alpha_a \beta_b   ~~~.\cr
}\eqno(14)$$

The only case involving significant work 
is the bracket of ${\bf Q}_{\alpha}$ with ${\bf Q}_{\beta}$, which on 
substituting the operator derivatives of the ${\bf Q}$'s into Eq.~(13), but 
before further algebraic rearrangement, takes the form 
$$\{ {\bf Q}_{\alpha} , {\bf Q}_{\beta} \}=
{\rm Tr} U~~~,\eqno(15a)$$
with $U$ given by 
$$\eqalign{
U=&2[X^j,\alpha_b 2 i \theta_a {1\over 2} \gamma_{ij}^{ab}]
[\beta_d, i\theta_c] i \gamma_i^{cd}  \cr
+&[X^m,[2i\theta_c,\beta_d]]{1\over 2}\gamma_{im}^{cd} 
 \alpha_b 2i \theta_a i \gamma_i^{ab} \cr
 -&i \big( i \gamma_i^{ab} p_{X_i} + {1\over 2} \gamma_{ij}^{ab}[X^i,X^j] 
 \big)
 \alpha_b {1 \over 2} \{\beta_d, i \gamma_{\ell}^{ad} p_{X_{\ell}}
 +{1 \over 2} \gamma_{\ell m}^{ad} [X^{\ell}, X^m] \} \cr
 +&{1\over 2}i \{\beta_d, i\gamma_{\ell}^{ad} p_{X_{\ell}}+{1\over 2}
 \gamma_{\ell m}^{ad} [X^{\ell},X^m] \} \big( i \gamma_i^{ab} p_{X_i} 
 +{1 \over 2} \gamma_{ij}^{ab}[X^i,X^j] \big) \alpha_b ~~~.\cr
 }\eqno(15b)$$
Algebraic rearrangement of this using the cyclic and gamma matrix identities 
gives ${\rm Tr}U={\rm Tr}T$,  
$$T=T_1 4i \delta^{bd}\alpha_b \beta_d 
+T_2^j 2 \gamma_j^{bd} \alpha_b \beta_d 
+X^{[\ell}X^mX^iX^{j]} 2i (\gamma_{[\ell}\gamma_m\gamma_i\gamma_{j]})^{bd}
\alpha_b \beta_d~~~,\eqno(16a)$$
with 
$$\eqalign{
T_1=&{1\over 2}p_{X^{\ell}}p_{X_{\ell}} -{1\over 4} [X_{\ell},X_m]
[X^{\ell},X^m]+{1 \over 2} [\theta_a,[X^j,\theta_c]]\gamma_j^{ac}  \cr 
{\rm Tr} T_1=& {\bf H} \cr 
}\eqno(16b)$$
and 
$$\eqalign{
T_2^j=&\{p_{X^i},[X^i,X^j]\}-i[\theta_a,[X^j,\theta^a]]  \cr
=&-\{X^j,\tilde C\} +[X^i,\{p_{X^i},X^j\}]-i\{\theta_a,\{X^j,\theta^a\}\} 
~~~, \cr
}\eqno(16c)$$
where in getting the final line of Eq.~(16c) we have used the 
Jacobi identities for mixed commutators and anticommutators.  
Equations (14) and (16a-c) 
agree with the charge density algebra computed in Ref. [3].  In the case 
when $\beta$ is a $c$-number, we see that the generalized Poisson 
bracket of the trace supercharges gives the trace Hamiltonian, showing that 
we have constructed a supersymmetry representation over noncommutative 
trace class operator dynamical variables.  As noted earlier, we will show 
elsewhere [8] that this construction works for general rigid supersymmetric 
theories, and in particular for the Wess-Zumino and supersymmetric Yang-Mills 
theories in four dimensions.  

We close with a remark on the quantization of the matrix model just 
described.  There are two possible points of view one could take.  The 
first possibility would be to quantize by replacing the Poisson brackets for 
matrix elements used in the computations of Ref. [3] 
by a commutator/anticommutator algebra; in this 
case one would be treating the matrix elements themselves as quantum 
operators.  An alternative possibility is discussed in Ref. [6], which is 
to regard the trace dynamics as fundamental without quantization, 
and to consider its statistical mechanics assuming ergodicity 
(an assumption that may presuppose  
taking the large $N$ limit; we also remark that the proof of the generalized 
Liouville theorem in Ref. [6] extends, with minor modifications, to the case   

employed here in which the fermions are realized with Grassmann matrices.)  
It is then shown that the statistical 
averages of the dynamical variables obey the rules of complex quantum field 
theory, with the effective Planck constant given by the expectation of the 
conserved operator $\tilde C$ discovered by Millard [7], which for a generic 
trace dynamics  model has the form
$$\tilde C=\sum_{\rm bosons}[q_i,p_i]-\sum_{\rm fermions}\{q_i,p_i\}   
~~~.\eqno(17a)$$
Since for the trace Lagrangian of Eq.~(3) $\tilde C$ is just the expression 
of Eq.~(7b), which vanishes as a constraint, the analysis of Ref. [6] 
implies that the statistical averages will obey {\it classical mechanics}, 
with the $X_i$ all commuting with one another and with 
the corresponding $p_{X_j}$.  As discussed in Ref. [4], it 
is easy to modify the model so that the constraint
of Eq.~(7b) reads instead $\tilde C=i \hbar$; this is done by adding to 
{\bf L} the term
$$\Delta {\bf L}= -{\rm Tr} A_0  ~~~,\eqno(17b)$$
which gives a trace action $\Delta {\bf S}=\int dt {\bf L}$ that is 
invariant under gauge transformations that vanish (or are periodic) at 
$t=\pm \infty$.  As we shall show in Ref. [8], it is easy to construct 
supersymmetric trace dynamics theories in which the operator 
$\tilde C$, although conserved, is not constrained to vanish; the  
trace dynamics extension of the Wess-Zumino model is an example.  

\bigskip
\centerline{\bf Acknowledgments}
This work was supported in part by the Department of Energy under
Grant \#DE--FG02--90ER40542.  I wish to thank Ed Witten, Andrew Millard, 
and members of 
the Princeton graduate student supersymmetry discussion group, for useful 
conversations.  
\vfill\eject
\centerline{\bf References}
\bigskip
\noindent
\item{[1]}  T. Banks, W. Fischler, S. H. Shenker, and L. Susskind, 
``M Theory as A Matrix Model: A Conjecture'', hep-th/9610043.
\bigskip 
\noindent
\item{[2]}  M. Claudson and M. B. Halpern, Nucl. Phys. B250 (1985) 689; 
V. Rittenberg and S. Yankielowicz, Ann. Phys. 162 (1985) 273; R. Flume, 
Ann. Phys. 164 (1985) 189.  For a recent survey, see B. de Wit, 
``Supersymmetric quantum mechanics, supermembranes and Dirichlet 
particles'', hep-th/9701169.
\bigskip
\noindent
\item{[3]}  T. Banks, N. Seiberg, and S. Shenker, ``Branes from 
Matrices'', hep-th/9612157.
\bigskip
\noindent
\item{[4]} S. L. Adler, Nucl. Phys. B 415 (1994) 195; S. L. Adler, 
``Quaternionic  Quantum Mechanics and Quantum Fields'', 
Sects. 13.5-13.7 and App. A (Oxford Univ. Press, New York, 1995).
\bigskip
\noindent
\item{[5]}  S. L. Adler, G. V. Bhanot, and J. D. Weckel, J. Math. Phys. 
35 (1994), 531; S. L. Adler and Y.-S. Wu, Phys. Rev. D 49 (1994) 6705.
\bigskip
\noindent
\item{[6]}  S. L. Adler and A. C. Millard, Nucl. Phys. B 473 (1996) 199. 
\bigskip
\noindent
\item{[7]}  A. C. Millard, Princeton University PhD thesis (in preparation).
\bigskip
\noindent
\item{[8]}  S. L. Adler, ``Poincar\'e Supersymmetry Representations 
Over Trace Class Noncommutative Graded Operator Algebras'', 
IASSNS-HEP-97/16 (in preparation).
\bigskip
\noindent
\item{[9]}  U. H. Danielsson, G. Ferretti, and B. Sundborg, ``D-Particle 
Dynamics and Bound States'', hep-th/9603081.
\bigskip
\noindent
\item{[10]} M. B. Green, J. H. Schwarz, and E. Witten,``Superstring 
Theory'', Vol. 1, p.246 (Cambridge Univ. Press, Cambridge, 1987).
\bigskip
\bye